# An extended class of $L^2$-series solutions of the wave equation

A. D. Alhaidari

*Physics Department, King Fahd University of Petroleum & Minerals, Dhahran 31261, Saudi Arabia*
e-mail: haidari@mailaps.org

We lift the constraint of a diagonal representation of the Hamiltonian by searching for square integrable bases that support an infinite tridiagonal matrix representation of the wave operator. The class of solutions obtained as such includes the discrete (for bound states) as well as the continuous (for scattering states) spectrum of the Hamiltonian. The problem translates into finding solutions of the resulting three-term recursion relation for the expansion coefficients of the wavefunction. These are written in terms of orthogonal polynomials, some of which are modified versions of known polynomials. The examples given, which are not exhaustive, include problems in one and three dimensions.

## I. INTRODUCTION

One of the advantages of obtaining exact solutions of the wave equation is that the analysis of such solutions makes the conceptual understanding of physics straightforward and sometimes intuitive. Moreover, these solutions are valuable means for checking and improving models and numerical methods introduced for solving complicated physical systems. In fact, in some limiting cases or for some special circumstances they may constitute analytic solutions of realistic problems or approximations thereof. Most of the known exactly solvable problems fall within distinct classes of, what is referred to as, "shape invariant potentials" [1]. Each class carries a representation of a given symmetry group. Supersymmetric quantum mechanics, potential algebras, point canonical transformations, and path integration are four methods among many which are used in the search for exact solutions of the wave equation. In nonrelativistic quantum mechanics, these developments were carried out over the years by many authors where several classes of these solutions are accounted for and tabulated (see, for example, the references cited in [1]). These formulations were extended to other classes of "conditionally exactly" [2,3] and "quasi exactly" [4,5] solvable problems where all or, respectively, part of the energy spectrum is known. Recently, the relativistic extension of some of these findings was carried out where several relativistic problems are formulated and solved exactly. These include, but not limited to, the Dirac-Morse, Dirac-Scarf, Dirac-Pöschl-Teller, Dirac-Hulthén...etc. [6].

In these developments, the main objective is to find solutions of the eigenvalue wave equation $H|\psi\rangle = E|\psi\rangle$, where $H$ is the Hamiltonian and $E$ is the energy which is either discrete (for bound states) or continuous (for scattering states). In most cases, especially when searching for algebraic or numerical solutions, the wave function $\psi$ spans the space of square integrable functions with discrete basis elements $\{\phi_n\}_{n=0}^{\infty}$. That is, the wavefunction is expandable as $|\psi(\vec{r}, E)\rangle = \sum_n f_n(E)|\phi_n(\vec{r})\rangle$, where $\vec{r}$ is the set of coordinates of real space. The basis functions must be compatible with the domain of the Hamiltonian. They should also satisfy the boundary conditions. Typically (and especially when calculating the discrete spectrum) the choice of basis is limited to those that carry a

diagonal representation of the Hamiltonian. That is, one looks for an $L^2$ basis set $\{\phi_n\}_{n=0}^{\infty}$ such that $H|\phi_n\rangle = E_n|\phi_n\rangle$ giving the discrete spectrum of $H$. The continuous spectrum is obtained from the analysis of an infinite sum of these *complete* basis functions. Truncating this sum, for numerical reasons, may create problems such as the presence of unphysical states or fictitious resonances in the spectrum.

In this article we relax the restriction of a diagonal matrix representation of the Hamiltonian. We only require that the hermitian matrix representation of the wave operator be tridiagonal. That is, the action of the wave operator on the elements of the basis is allowed to take the general form $(H-E)|\phi_n\rangle \sim |\phi_n\rangle + |\phi_{n-1}\rangle + |\phi_{n+1}\rangle$ and such that

$$\langle \phi_n | H - E | \phi_m \rangle = (a_n - z)\delta_{n,m} + b_n \delta_{n,m-1} + b_{n-1}\delta_{n,m+1}, \tag{1.1}$$

where $z$ and the coefficients $\{a_n, b_n\}_{n=0}^{\infty}$ are real and, in general, functions of the energy, angular momentum, and potential parameters. Therefore, the matrix wave equation, which is obtained by expanding $|\psi\rangle$ as $\sum_m f_m |\phi_m\rangle$ in $(H-E)|\psi\rangle = 0$ and projecting on the left by $\langle \phi_n |$, results in the following three-term recursion relation

$$z f_n = a_n f_n + b_{n-1} f_{n-1} + b_n f_{n+1}. \tag{1.2}$$

Consequently, the problem translates into finding solutions of the recursion relation for the expansion coefficients of the wavefunction $\psi$. In most cases this recursion is solved easily and directly by correspondence with those for well known orthogonal polynomials. It is obvious that the solution of (1.2) is obtained modulo an overall factor which is a function of the physical parameters of the problem but, otherwise, independent of $n$. The uniqueness of the solution is achieved by the requirement (for example) of normalizability of the wavefunction, that is $\langle \psi | \psi \rangle = 1$. It should also be noted that the solution of the problem as depicted by Eq. (1.2) above is obtained for all $E$, the discrete as well as the continuous. Moreover, the representation equation (1.1) clearly shows that the discrete spectrum is easily obtained by diagonalization which requires that:

$$b_n = 0, \text{ and } a_n - z = 0. \tag{1.3}$$

One could obtain two solutions to the recursion (1.2) by starting with two different initial relations ($n = 0$). One solution is associated with the homogeneous recursion relation that starts with:

$$(a_0 - z)f_0 + b_0 f_1 = 0. \tag{1.4}$$

The other is associated with a non-homogeneous recursion whose initial relation is

$$(a_0 - z)f_0 + b_0 f_1 = \xi, \tag{1.5}$$

where $\xi$ is a real and non-zero "seed" parameter. One of these two solutions behaves asymptotically ($n \to \infty$) as sine-like while the other behaves as cosine-like. These two solutions have the same asymptotic limits as the regular and irregular solutions of the second order differential wave equation. Scattering states and the phase shift could be obtained algebraically by studying these asymptotic limits. These are issues of concern in algebraic scattering theories such as the *J*-matrix method [7]. In the present work, however, we will only be concerned with the regular solutions of the wave equation.

In configuration space, with coordinate $x$, the wavefunction $\psi_E(x)$ is expanded as $\sum_{n=0}^{\infty} f_n(E)\phi_n(x)$ where the $L^2$ basis functions could generally be written as

$$\phi_n(x) = A_n w_n(x) P_n(x). \tag{1.6}$$

$A_n$ is a normalization constant, $P_n(x)$ is a polynomial of degree $n$ in $x$, and the weight function satisfies $w_n(x_\pm) = 0$, where $x_-(x_+)$ is the left (right) boundary of configuration space. In the following sections we consider examples in two spaces. One is where $x_\pm$ are finite and

$$w_n(x) = (x - x_-)^\alpha (x - x_+)^\beta,$$
$$P_n(x) = {}_2F_1(-n, b; c; x).$$
(1.7)

The other is semi-infinite where $x_-$ is finite, $x_+$ infinite, and where

$$w_n(x) = (x - x_-)^\alpha e^{-\beta(x - x_-)},$$
$$P_n(x) = {}_1F_1(-n; c; x).$$
(1.8)

${}_2F_1$ is the hypergeometric function and ${}_1F_1$ is the confluent hypergeometric function. The parameters $\alpha, \beta, b$ and $c$ are real with $\alpha$ and $\beta$ positive. They are related to the physical parameters of the corresponding problem and may also depend (for bound states) on the index $n$.

In the following sections we consider examples of various problems in one and three dimensions and find their $L^2$ series solutions. The solutions of some of the classic problems such as the Coulomb and Morse are reproduced adding, however, new tridiagonal representations of the solution space. We also find generalizations of others such as the Hulthén-type problems where we obtain an extended class of solutions and define their associated orthogonal polynomials. In addition, we investigate problems with hyperbolic potentials such as the Rosen-Morse type and present its generalized solution. These investigations do not exhaust the set of all solvable problems using this approach. Furthermore, this development embodies powerful tools in the analysis of solutions of the wave equation by exploiting the intimate connection and interplay between tridiagonal matrices and the theory of orthogonal polynomials. In such analysis, one is at liberty to employ a wide range of well established methods and numerical techniques associated with these settings such as quadrature approximation and continued fractions [8].

## II. THE COULOMB PROBLEM

We start by taking the configuration space coordinate simply as $x = \lambda r$, where $\lambda$ is a length scale parameter which is real and positive and $r$ is the radial coordinate in three dimensions. This problem belongs to the case described by Eq. (1.8) with $x_- = 0$. Since ${}_1F_1(-n; c; z)$ is proportional to the Laguerre polynomial $L_n^{c-1}(z)$ [9], we could write the basis functions as

$$\phi_n(r) = A_n x^\alpha e^{-\beta x} L_n^\nu(x),$$
(2.1)

where $A_n = \sqrt{\lambda \Gamma(n+1)/\Gamma(n+\nu+1)}$, $n = 0, 1, 2, \ldots$, $\nu > -1$, $\alpha$ and $\beta$ are real and positive. In the atomic units, $\hbar = m = 1$, the radial time-independent Schrödinger wave equation for a structureless particle in a spherically symmetric potential $V(r)$ reads

$$(H - E)|\psi\rangle = \left[ -\frac{1}{2} \frac{d^2}{dr^2} + \frac{\ell(\ell+1)}{2r^2} + V(r) - E \right] |\psi\rangle = 0,$$
(2.2)

where $\ell$ is the angular momentum quantum number. The action of the first term (the second order derivative) on the basis function results in the following

$$\frac{d^2\phi_n}{dr^2} = \lambda^2 A_n x^\alpha e^{-\beta x}\left[\frac{d^2}{dx^2} + \left(\frac{2\alpha}{x} - 2\beta\right)\frac{d}{dx} - \frac{\alpha}{x^2} + \left(\frac{\alpha}{x} - \beta\right)^2\right]L_n^\nu(x). \quad (2.3)$$

Using the differential equation and differential formula of the Laguerre polynomial [Eqs. (A.3) and (A.4) in the Appendix] we obtain the following action of the wave operator (2.2) on the basis

$$\frac{2}{\lambda^2}(H-E)|\phi_n\rangle = \left[\frac{n}{x}\left(2\beta + \frac{\nu+1-2\alpha}{x}\right) + \frac{\ell(\ell+1)-\alpha(\alpha-1)}{x^2} + \frac{2\alpha\beta}{x} - \beta^2 \right.$$
$$\left. + \frac{2}{\lambda^2}(V-E)\right]|\phi_n\rangle + \frac{n+\nu}{x}\left(1-2\beta + \frac{2\alpha-\nu-1}{x}\right)\frac{A_n}{A_{n-1}}|\phi_{n-1}\rangle. \quad (2.4)$$

The orthogonality relation for the Laguerre polynomials [shown in the Appendix as Eq. (A.5)] requires that $\beta = 1/2$ if we were to obtain a tridiagonal representation for $\langle\phi_n|H-E|\phi_m\rangle$. Additionally, we end up with only two possibilities to achieve the tridiagonal structure of the wave operator (2.4):

(1) $\nu = 2\alpha - 1$, $\alpha = \ell + 1$, and $V = \frac{\mathcal{Z}}{r}$ \hfill (2.5a)

(2) $\nu = 2\alpha - 2$, $\lambda^2 = -8E$, and $V = \frac{\mathcal{Z}}{r} + \frac{B/2}{r^2}$ \hfill (2.5b)

where $\mathcal{Z}$ and $B$ are real potential parameters. $\mathcal{Z}$ is the particle's charge and $B$ is a centripetal potential barrier parameter.

We start by considering the first possibility described by (2.5a) where we have:
$$\phi_n(r) = A_n x^{\ell+1} e^{-x/2} L_n^{2\ell+1}(x). \quad (2.6)$$

Substituting in Eq. (2.4) with $\beta = 1/2$ and projecting on $\langle\phi_m|$ we obtain

$$\frac{2}{\lambda^2}\langle\phi_m|H-E|\phi_n\rangle = \left[2(n+\ell+1)\left(\tfrac{1}{4} - \tfrac{2E}{\lambda^2}\right) + \tfrac{2\mathcal{Z}}{\lambda}\right]\delta_{n,m}$$
$$+ \left(\tfrac{1}{4} + \tfrac{2E}{\lambda^2}\right)\left[\sqrt{n(n+2\ell+1)}\delta_{n,m+1} + \sqrt{(n+1)(n+2\ell+2)}\delta_{n,m-1}\right]. \quad (2.7)$$

Therefore, the resulting recursion relation (1.2) for the expansion coefficients of the wavefunction becomes

$$\left[2(n+\ell+1)\frac{\sigma_-}{\sigma_+} - \frac{2\mathcal{Z}/\lambda}{\sigma_+}\right]f_n - \sqrt{n(n+2\ell+1)}f_{n-1} - \sqrt{(n+1)(n+2\ell+2)}f_{n+1} = 0, \quad (2.8)$$

where $\sigma_\pm = \frac{2E}{\lambda^2} \pm \frac{1}{4}$. Rewriting this recursion in terms of the polynomials $P_n(E) = \sqrt{\Gamma(n+2\ell+2)/\Gamma(n+1)}\, f_n(E)$, we obtain the more familiar recursion relation

$$2\left[(n+\ell+1)\frac{\sigma_-}{\sigma_+} - \frac{\mathcal{Z}/\lambda}{\sigma_+}\right]P_n - (n+2\ell+1)P_{n-1} - (n+1)P_{n+1} = 0, \quad (2.9)$$

which is that of the Pollaczek polynomials [10] provided that $E > 0$ [see Eq. (A.11) in the Appendix]. Thus, we can write

$$f_n(E) = \sqrt{\frac{\Gamma(n+1)}{\Gamma(n+2\ell+2)}}\, P_n^{\ell+1}\left(\frac{-\mathcal{Z}}{\sqrt{2E}}, \cos^{-1}\frac{E-\lambda^2/8}{E+\lambda^2/8}\right), \quad (2.10)$$

giving the following $L^2$-series solution of the Coulomb problem for positive energies

$$\psi^\ell(r,E) = N^\ell \sum_{n=0}^{\infty} \frac{\Gamma(n+1)}{\Gamma(n+2\ell+2)} P_n^{\ell+1}\left(\frac{-\mathcal{Z}}{\sqrt{2E}}, \cos^{-1}\frac{E-\lambda^2/8}{E+\lambda^2/8}\right)(\lambda r)^{\ell+1} e^{-\lambda r/2} L_n^{2\ell+1}(\lambda r), \quad (2.11)$$

where $N^\ell$ is a normalization constant. This solution of the Coulomb problem was obtained by Yamani and Reinhardt [11]. Restricting the representation (2.7) to the diagonal form gives the discrete spectrum via the requirement (1.3) which, in this case, reads as follows:

$$\tfrac{1}{4} + \tfrac{2E}{\lambda^2} = 0,$$
$$2(n+\ell+1)\left(\tfrac{1}{4} - \tfrac{2E}{\lambda^2}\right) + \tfrac{2\mathcal{Z}}{\lambda} = 0 \, .$$
(2.12)

This gives the following well known energy spectrum for the bound states of the Coulomb problem

$$E_n = -\tfrac{1}{2}\left(\tfrac{\mathcal{Z}}{n+\ell+1}\right)^2, \quad \lambda = \lambda_n = 2|\mathcal{Z}|/(n+\ell+1),$$
(2.13)

where $n = 0,1,2,...$. Therefore, the corresponding bound states wavefunctions are

$$\psi_n^\ell(r) = A_n (\lambda_n r)^{\ell+1} e^{-\lambda_n r/2} L_n^{2\ell+1}(\lambda_n r) \, .$$
(2.14)

For the second possibility described by (2.5b) the basis functions are

$$\phi_n(r) = A_n x^{1+\nu/2} e^{-x/2} L_n^\nu(x), \quad x = 2\sqrt{-2E}\, r \, ,$$
(2.15)

where $E < 0$. Thus, the basis parameter in this case is the Laguerre polynomial index $\nu$. Substituting in Eq. (2.4) and projecting on $\langle\phi_m|$ we get

$$-\tfrac{1}{4E}\langle\phi_m|H - E|\phi_n\rangle = \left[(2n+\nu+1)\left(n+\tfrac{\nu}{2}+1+\tau\right) - n - \left(\tfrac{\nu+1}{2}\right)^2 + (\ell+\tfrac{1}{2})^2 + B\right]\delta_{n,m}$$
$$-\left(n+\tfrac{\nu}{2}+\tau\right)\sqrt{n(n+\nu)}\delta_{n,m+1} - \left(n+\tfrac{\nu}{2}+1+\tau\right)\sqrt{(n+1)(n+\nu+1)}\delta_{n,m-1}$$
(2.16)

where $\tau = \mathcal{Z}/\sqrt{-2E}$. The resulting recursion relation for the expansion coefficients of the wavefunction in terms of the polynomials $P_n(E) = \sqrt{\Gamma(n+1)/\Gamma(n+\nu+1)}\, f_n(E)$ reads

$$\left[(2n+\nu+1)\left(n+\tfrac{\nu}{2}+1+\tau\right) - n - \left(\tfrac{\nu+1}{2}\right)^2 + (\ell+\tfrac{1}{2})^2 + B\right]P_n$$
$$-n\left(n+\tfrac{\nu}{2}+\tau\right)P_{n-1} - (n+\nu+1)\left(n+\tfrac{\nu}{2}+1+\tau\right)P_{n+1} = 0 \, .$$
(2.17)

This is a special case of the recursion relation for the continuous dual Hahn orthogonal polynomial [12], which is shown as Eq. (A.14) in the Appendix. As a result we obtain

$$f_n(E) = \sqrt{\tfrac{\Gamma(n+\nu+1)}{\Gamma(n+1)}} \times \begin{cases} S_n^{\frac{\nu+1}{2}}\left(z; \tfrac{\nu+1}{2}, \tau+\tfrac{1}{2}\right) &, B < -(\ell+\tfrac{1}{2})^2 \\ \tilde{S}_n^{\frac{\nu+1}{2}}\left(z; \tfrac{\nu+1}{2}, \tau+\tfrac{1}{2}\right) &, B > -(\ell+\tfrac{1}{2})^2 \end{cases}$$
(2.18)

where $z = \sqrt{|(\ell+\tfrac{1}{2})^2 + B|}$ and $\tilde{S}_n^\mu(z;a,b)$ is a modified version of the continuous dual Hahn polynomial defined as

$$\tilde{S}_n^\mu(x;a,b) \equiv S_n^\mu(-ix;a,b) = {}_3F_2\left(\begin{matrix}-n, \mu+x, \mu-x\\ \mu+a, \mu+b\end{matrix}\bigg|1\right)$$
(2.19)

The discrete energy spectrum is evidently obtained by diagonalization of (2.16) which translates into the requirements

$$n + \tfrac{\nu}{2} + 1 + \tau = 0,$$
$$-\left(\tfrac{\nu+1}{2}\right)^2 + (\ell+\tfrac{1}{2})^2 + B = 0,$$
(2.20)

giving,

$$E_n = -\tfrac{1}{2}\left(\tfrac{\mathcal{Z}}{n+\nu/2+1}\right)^2, \quad \nu = -1 + 2\sqrt{|(\ell+\tfrac{1}{2})^2 + B|} \, .$$
(2.21)

This is the same energy spectrum as that in Eq. (2.13) above when $B = 0$.

## III. THE SPHERICAL OSCILLATOR

In this case, we write the configuration space coordinate as $x = (\lambda r)^2$. Thus, the basis elements of the $L^2$ space is written as

$$\phi_n(r) = A_n x^\alpha e^{-\beta x} L_n^\nu(x), \tag{3.1}$$

where again $\alpha$ and $\beta$ are real and positive, $\nu > -1$, and $A_n = \sqrt{2|\lambda|\Gamma(n+1)/\Gamma(n+\nu+1)}$. Using $\frac{d}{dr} = 2|\lambda|\sqrt{x}\frac{d}{dx}$, we obtain the following

$$\frac{d^2\phi_n}{dr^2} = 4\lambda^2 A_n x^{\alpha+1} e^{-\beta x} \left[ \frac{d^2}{dx^2} + \left(\tfrac{2\alpha+\frac{1}{2}}{x} - 2\beta\right)\frac{d}{dx} - \frac{\alpha}{x^2} + \left(\tfrac{\alpha+\frac{1}{2}}{x} - \beta\right)\left(\tfrac{\alpha}{x} - \beta\right) \right] L_n^\nu(x). \tag{3.2}$$

Employing the differential formulas of the Laguerre polynomials, which are shown in the Appendix, gives the following

$$\frac{d^2\phi_n}{dr^2} = 4\lambda^2 x \left[ \frac{n}{x}\left(\tfrac{2\alpha-\nu-\frac{1}{2}}{x} - 2\beta\right) + \frac{\alpha(\alpha-\frac{1}{2})}{x^2} - \frac{\beta(2\alpha+\frac{1}{2})}{x} + \beta^2 \right]\phi_n$$
$$- 4\lambda^2 (n+\nu)\left(\tfrac{2\alpha-\nu-\frac{1}{2}}{x} + 1 - 2\beta\right)\frac{A_n}{A_{n-1}}\phi_{n-1}. \tag{3.3}$$

The tridiagonal structure can only be achieved if and only if $\beta = \frac{1}{2}$ resulting in the following action of the wave operator on the basis

$$\frac{1}{2\lambda^2}(H-E)|\phi_n\rangle = \left[ -\frac{n(2\alpha-\nu-\frac{1}{2}) + (\alpha-\frac{1}{4})^2 - (\ell+\frac{1}{2})^2/4}{x} + n + \alpha + \frac{1}{4} - \frac{x}{4} \right.$$
$$\left. + \frac{1}{2\lambda^2}(V-E)\right]|\phi_n\rangle + \frac{(n+\nu)(2\alpha-\nu-\frac{1}{2})}{x}\frac{A_n}{A_{n-1}}|\phi_{n-1}\rangle. \tag{3.4}$$

The orthogonality relation (A.5) and the tridiagonal requirement limit the possibilities to either one of the following two:

(1) $\nu = 2\alpha - 1/2$, $2\alpha = \ell + 1$, and $V = \frac{1}{2}\omega^4 r^2$ \hfill (3.5a)

(2) $\nu = 2\alpha - 3/2$, and $V = \frac{1}{2}\lambda^4 r^2 + B/2r^2$ \hfill (3.5b)

where $\omega$ is the oscillator frequency and $B$ is a centripetal potential barrier parameter.

The first possibility gives the following tridiagonal matrix representation of the wave operator

$$\frac{2}{\lambda^2}\langle\phi_m|H-E|\phi_n\rangle = \left[(2n+\nu+1)\left(\tfrac{\omega^4}{\lambda^4}+1\right) - \tfrac{2E}{\lambda^2}\right]\delta_{n,m}$$
$$- \left(\tfrac{\omega^4}{\lambda^4}-1\right)\left[\sqrt{n(n+\nu)}\delta_{n,m+1} + \sqrt{(n+1)(n+\nu+1)}\delta_{n,m-1}\right]. \tag{3.6}$$

Writing the resulting recursion relation in terms of the polynomials $P_n(E) = \sqrt{\Gamma(n+\nu+1)/\Gamma(n+1)}\, f_n(E)$ gives the following

$$2\left[\left(n+\tfrac{\nu+1}{2}\right)\tfrac{\sigma_+}{\sigma_-} - \tfrac{E/\lambda^2}{\sigma_-}\right]P_n - (n+\nu)P_{n-1} - (n+1)P_{n+1} = 0, \tag{3.7}$$

where $\sigma_\pm = (\omega/\lambda)^4 \pm 1$. We compare this with the recursion relation of the Pollaczek polynomial [Eq. (A.11) in the Appendix] while carefully considering the range of values of the parameters. Using the well known relations, $\cosh\theta = \cos i\theta$ and $\sinh\theta = -i\sin i\theta$, we could define a "Hyperbolic Pollaczek polynomial" as follows

$$\tilde{P}_n^\mu(x,\theta) \equiv P_n^\mu(-ix,i\theta) = \frac{\Gamma(n+2\mu)}{\Gamma(n+1)\Gamma(2\mu)} e^{-n\theta} {}_2F_1(-n,\mu+x;2\mu;1-e^{2\theta}), \tag{3.8}$$

where $\theta > 0$. These polynomials satisfy the following modified three-term recursion relation:

$$2[(n+\mu)\cosh\theta + x\sinh\theta]\tilde{P}_n^\mu - (n+2\mu-1)\tilde{P}_{n-1}^\mu - (n+1)\tilde{P}_{n+1}^\mu = 0 \tag{3.9}$$

Therefore, the expansion coefficients of the oscillator wavefunction could be written in terms of these polynomials and in one of two alternative expressions, depending on the values of the parameters, as follows:

$$f_n(E) = \sqrt{\frac{\Gamma(n+1)}{\Gamma(n+\ell+3/2)}} \times \begin{cases} \tilde{P}_n^{\frac{\ell+3/2}{2}}\left(\frac{-E}{2\omega^2}, \sinh^{-1}\frac{2\rho^2}{\rho^4-1}\right) & , |\lambda| < \omega \\ (-)^n \tilde{P}_n^{\frac{\ell+3/2}{2}}\left(\frac{-E}{2\omega^2}, \sinh^{-1}\frac{2\rho^2}{1-\rho^4}\right) & , |\lambda| > \omega \end{cases} \tag{3.10}$$

where $\rho = \omega/\lambda$. The discrete energy spectrum is easily obtained by diagonalization of the Hamiltonian in (3.6) resulting in the following requirements

$$\lambda^2 = \omega^2, \text{ and } 2n+\nu+1 = E/\lambda^2, \tag{3.11}$$

which gives

$$E_n = \omega^2(2n+\nu+1), \tag{3.12}$$

where $\nu = \ell + 1/2$ and $n = 0, 1, 2, \ldots$. Thus, the bound states wavefunctions are

$$\psi_n^\ell(r) = A_n(\omega r)^{\ell+1} e^{-\omega^2 r^2/2} L_n^{\ell+1/2}(\omega^2 r^2). \tag{3.13}$$

Now, we consider the second possibility (3.5b) in which the parameter $\nu$, aside from being $> -1$, is arbitrary. It is to be noted that in this case the first term in the potential, $\frac{1}{2}\lambda^2 x$, is essential to cancel the contribution of the term $-\frac{x}{4}$ in Eq. (3.4) which destroys the tridiagonal structure. Therefore, the basis scale parameter should be identified with the oscillator frequency, whereas, the arbitrary basis parameter is the Laguerre polynomial index $\nu$. Following the same procedure as outlined above we obtain the following matrix representation of the wave operator

$$\frac{1}{2\lambda^2}\langle\phi_m|H-E|\phi_n\rangle = \left[(2n+\nu+1)\left(n+\tfrac{\nu}{2}+1+\tau\right) - n - \left(\tfrac{\nu+1}{2}\right)^2 + \left(\tfrac{\ell+\frac{1}{2}}{2}\right)^2 + \tfrac{B}{4}\right]\delta_{n,m}$$
$$-\left(n+\tfrac{\nu}{2}+\tau\right)\sqrt{n(n+\nu)}\delta_{n,m+1} - \left(n+\tfrac{\nu}{2}+1+\tau\right)\sqrt{(n+1)(n+\nu+1)}\delta_{n,m-1}, \tag{3.14}$$

where $\tau = -E/2\lambda^2$. Similarly, one could easily show that the solution of the recursion relation (1.2) resulting from the matrix wave equation is written in terms of the continuous dual Hahn orthogonal polynomial and its "modified" version as follows:

$$f_n(E) = \sqrt{\frac{\Gamma(n+\nu+1)}{\Gamma(n+1)}} \times \begin{cases} S_n^{\frac{\nu+1}{2}}\left(z;\frac{\nu+1}{2},\tau+\frac{1}{2}\right) & , B < -(\ell+\tfrac{1}{2})^2 \\ \tilde{S}_n^{\frac{\nu+1}{2}}\left(z;\frac{\nu+1}{2},\tau+\frac{1}{2}\right) & , B > -(\ell+\tfrac{1}{2})^2 \end{cases} \tag{3.15}$$

where $z = \tfrac{1}{2}\sqrt{|(\ell+\tfrac{1}{2})^2 + B|}$ and $\tilde{S}_n^\mu(x;a,b)$ is defined by Eq. (2.19). Additionally, the discrete energy spectrum is obtained from (3.14) by the requirement that

$$n + \tfrac{\nu}{2} + 1 + \tau = 0,$$
$$-\left(\tfrac{\nu+1}{2}\right)^2 + \left(\tfrac{\ell+\frac{1}{2}}{2}\right)^2 + \tfrac{B}{4} = 0. \tag{3.16}$$

Giving:

$$E_n = \lambda^2(2n+\nu+2), \quad \nu = -1 + \sqrt{|(\ell+\tfrac{1}{2})^2 + B|}, \tag{3.17}$$

which is the same energy spectrum as that in Eq. (3.12) when $B = 0$.

## IV. POWER-LAW POTENTIALS AT ZERO ENERGY

The configuration space coordinate for this problem is $x = (\lambda r)^\gamma$, where $\lambda > 0$ and the real parameter $\gamma \neq 0, 1, 2$. The three dismissed values of $\gamma$ correspond to the Morse, Coulomb, and Oscillator problems, respectively [13]. An element of the basis functions could be taken as follows
$$\phi_n(r) = A_n x^\alpha e^{-x/2} L_n^\nu(x). \tag{4.1}$$
$\phi_n(0) = \phi_n(\infty) = 0$ for positive and negative values of $\gamma$, which is fixed once and for all. The integration measure in terms of $x$ is $\frac{1}{\lambda|\gamma|} x^{-1+1/\gamma} dx$ since for $\pm\gamma > 0$ we get $\int_0^\infty dr = \frac{\pm 1}{\lambda\gamma} \int_0^\infty x^{-1+1/\gamma} dx$. Therefore, the normalization constant is $A_n = \sqrt{|\gamma|\lambda \Gamma(n+1)/\Gamma(n+\nu+1)}$. Using the derivative chain rule, which gives $\frac{d}{dr} = \gamma \lambda x^{1-1/\gamma} \frac{d}{dx}$, we can write

$$\frac{2}{(\gamma\lambda)^2}(H-E)|\phi_n\rangle = x^{-2/\gamma}\left[\left(\frac{\ell+\frac{1}{2}}{\gamma}\right)^2 - \left(\alpha - \frac{1}{2\gamma}\right)^2 - n\left(2\alpha - \nu - \frac{1}{\gamma}\right)\right.$$
$$\left. + x\left(n + \alpha + \frac{1}{2} - \frac{1}{2\gamma}\right) - \frac{x^2}{4} + \frac{2}{(\gamma\lambda)^2} x^{2/\gamma}(V-E)\right]|\phi_n\rangle \tag{4.2}$$
$$+ x^{-2/\gamma}(n+\nu)\left(2\alpha - \nu - \frac{1}{\gamma}\right)\frac{A_n}{A_{n-1}}|\phi_{n-1}\rangle.$$

The constant $E$ term must independently vanish (for $\gamma \neq 0, 1, 2$) if the representation of the wave operator is to be tridiagonal. Consequently, this case is analytically solvable only for zero energy. It is to be noted that this condition does not diminish the significance of these solutions. Zero energy solutions have valuable applications in scattering calculations (e.g., effective rang and scattering length parameters [14]) and in the investigation of low energy limit cases.

For this problem we also end up with two possibilities for obtaining the tridiagonal structure of the Hamiltonian:

(1) $\nu = 2\alpha - 1/\gamma$, $\alpha = \begin{cases} (\ell+1)/\gamma & ,\gamma > 0 \\ -\ell/\gamma & ,\gamma < 0 \end{cases}$, and $V = \frac{1}{r^2}\left(Ar^\gamma + \frac{1}{2}Br^{2\gamma}\right)$ (4.3a)

(2) $\nu = 2\alpha - 1 - 1/\gamma$, and $V = \frac{1}{r^2}\left[A + \frac{1}{2}Br^\gamma + \frac{1}{2}(\gamma/2)^2(\lambda r)^{2\gamma}\right]$ (4.3b)

where $A$ and $B$ are real potential parameters. The last term of the potential in the second case (4.3b) is necessary to eliminate the non-tridiagonal component coming from the contribution of the term $-\frac{1}{4}x^{2-2/\gamma}$ in Eq. (4.2).

The resulting representation of the Hamiltonian for the first case (4.3a) is
$$\frac{2}{\lambda^2}\langle\phi_m|H|\phi_n\rangle = \left[\left(\frac{B}{\lambda^{2\gamma}} + \frac{\gamma^2}{4}\right)(2n+\nu+1) + \frac{2A}{\lambda^\gamma}\right]\delta_{n,m}$$
$$- \left(\frac{B}{\lambda^{2\gamma}} - \frac{\gamma^2}{4}\right)\left[\sqrt{n(n+\nu)}\delta_{n,m+1} + \sqrt{(n+1)(n+\nu+1)}\delta_{n,m-1}\right]. \tag{4.4}$$

The recursion relation obtained from this representation has two solutions depending on whether $B$ is positive or negative. For $B < 0$ and $\pm\gamma > 0$, we get:

$$f_n(\ell) = \sqrt{\frac{\Gamma(n+1)}{\Gamma(n+1\pm\frac{2\ell+1}{\gamma})}} P_n^{\pm\frac{\ell+\frac{1}{2}}{\gamma}+\frac{1}{2}}\left(\frac{-A}{|\gamma|\sqrt{-B}}, \cos^{-1}\frac{\rho^2-1}{\rho^2+1}\right), \tag{4.5}$$

where $\rho = 2\sqrt{|B|}/|\gamma|\lambda^\gamma$. On the other hand, for $B > 0$ and $\pm\gamma > 0$ the solution is written in terms of the Hyperbolic Pollaczek polynomial as follows:

$$f_n(\ell) = \sqrt{\frac{\Gamma(n+1)}{\Gamma(n+1\pm\frac{2\ell+1}{\gamma})}} \times \begin{cases} \tilde{P}_n^{\pm\frac{\ell+\frac{1}{2}}{\gamma}+\frac{1}{2}}\left(\frac{A}{|\gamma|\sqrt{B}}, \sinh^{-1}\frac{2\rho}{\rho^2-1}\right) & ,\rho > 1 \\ (-)^n \tilde{P}_n^{\pm\frac{\ell+\frac{1}{2}}{\gamma}+\frac{1}{2}}\left(\frac{A}{|\gamma|\sqrt{B}}, \sinh^{-1}\frac{2\rho}{1-\rho^2}\right) & ,\rho < 1 \end{cases} \tag{4.6}$$

The diagonal representation of the Hamiltonian is obtained from Eq. (4.4) by the requirement that the potential parameters assume the following values:

$$B = (\gamma\lambda^\gamma/2)^2, \text{ and } A = -\lambda^\gamma(\gamma/2)^2\left(2n+1\pm\frac{2\ell+1}{\gamma}\right) \text{ for } \pm\gamma > 0 \tag{4.7}$$

Thus, the discrete spectrum of the Hamiltonian occurs for positive values of the potential parameter $B$ and for negative, $\ell$-dependent, and discrete values of $A$ with $\Delta A = \gamma^2 \lambda^\gamma/2$. The eigenfunctions of the Hamiltonian that correspond to this discrete representation and for $\pm\gamma > 0$ are as follows:

$$\psi_n^\ell(r) = A_n (\lambda r)^{\frac{1}{2}\pm(\ell+\frac{1}{2})} e^{-\lambda^\gamma r^\gamma/2} L_n^{\pm(2\ell+1)/\gamma}(\lambda^\gamma r^\gamma), \tag{4.8}$$

This diagonal representation has already been obtained by this author [13] and others [15].

The second possibility defined in (4.3b) produces the following tridiagonal matrix representation for $H$

$$\frac{2}{(\gamma\lambda)^2}\langle\phi_m|H|\phi_n\rangle = \left[(2n+\nu+1)\left(n+\frac{\nu}{2}+1+\tau\right) - n - \left(\frac{\nu+1}{2}\right)^2 + \left(\frac{\ell+\frac{1}{2}}{\gamma}\right)^2 + \frac{2A}{\gamma^2}\right]\delta_{n,m}$$
$$-\left(n+\frac{\nu}{2}+\tau\right)\sqrt{n(n+\nu)}\delta_{n,m+1} - \left(n+\frac{\nu}{2}+1+\tau\right)\sqrt{(n+1)(n+\nu+1)}\delta_{n,m-1} \tag{4.9}$$

where $\tau = B/\gamma^2\lambda^\gamma$ and $\nu > -1$ but, otherwise, arbitrary. The associated recursion relation is solved for the expansion coefficients of the wavefunction in terms of the continuous dual Hahn polynomial and its "modified" version as follows:

$$f_n(\ell) = \sqrt{\frac{\Gamma(n+\nu+1)}{\Gamma(n+1)}} \times \begin{cases} S_n^{\frac{\nu+1}{2}}\left(z; \frac{\nu+1}{2}, \tau+\frac{1}{2}\right) & ,2A < -(\ell+\frac{1}{2})^2 \\ \tilde{S}_n^{\frac{\nu+1}{2}}\left(z; \frac{\nu+1}{2}, \tau+\frac{1}{2}\right) & ,2A > -(\ell+\frac{1}{2})^2 \end{cases} \tag{4.10}$$

where $z = \frac{1}{|\gamma|}\sqrt{|(\ell+\frac{1}{2})^2 + 2A|}$. The diagonal representation is obtained by restricting the potential and basis parameters to satisfy

$$B = -\gamma^2 \lambda^\gamma (n+\nu/2+1),$$
$$2A = \left(\gamma\frac{\nu+1}{2}\right)^2 - (\ell+\frac{1}{2})^2, \tag{4.11}$$

which is the same results as that in (4.7) above with $A = 0$ and $B \to 2A$ as it is also evident by comparing the potential in (4.3a) with that in (4.3b).

## V. THE ONE-DIMENSIONAL MORSE OSCILLATOR

In this example, we only list the results without giving details of the calculation:

$$x = \mu e^{-\lambda y}, \text{ where } \mu, \lambda > 0 \text{ and } y \in \Re^1 \text{ (the real line)} \tag{5.1}$$

$$\phi_n(r) = A_n x^\alpha e^{-x/2} L_n^\nu(x), \quad A_n = \sqrt{\lambda \Gamma(n+1)/\Gamma(n+\nu+1)} \tag{5.2}$$

Case (1) $\nu = 2\alpha$:

$$V = \tfrac{\lambda^2}{2}\left(A e^{-\lambda r} + B e^{-2\lambda r}\right), \quad E = -\tfrac{1}{2}(\lambda\alpha)^2. \tag{5.3}$$

$$\frac{2}{\lambda^2}\langle \phi_m | H - E | \phi_n \rangle = \left[(2n+\nu+1)\left(\tfrac{B}{\mu^2}+\tfrac{1}{4}\right)+\tfrac{A}{\mu}\right]\delta_{n,m}$$
$$-\left(\tfrac{B}{\mu^2}-\tfrac{1}{4}\right)\left[\sqrt{n(n+\nu)}\,\delta_{n,m+1}+\sqrt{(n+1)(n+\nu+1)}\,\delta_{n,m-1}\right] \tag{5.4}$$

where $\nu = \tfrac{2}{\lambda}\sqrt{-2E}$ and $E < 0$. The solution for $B < 0$ is

$$f_n(E) = \sqrt{\tfrac{\Gamma(n+1)}{\Gamma(n+\nu+1)}} P_n^{\tfrac{\nu+1}{2}}\left(\tfrac{-A}{2\sqrt{-B}}, \cos^{-1}\tfrac{\rho^2-1}{\rho^2+1}\right), \tag{5.5}$$

where $\rho = 2\sqrt{|B|}/\mu$. However, if $B$ is positive then we obtain the following

$$f_n(E) = \sqrt{\tfrac{\Gamma(n+1)}{\Gamma(n+\nu+1)}} \times \begin{cases} \tilde{P}_n^{\tfrac{\nu+1}{2}}\left(\tfrac{A}{2\sqrt{B}}, \sinh^{-1}\tfrac{2\rho}{\rho^2-1}\right) & ,\mu < 2\sqrt{B} \\ (-)^n \tilde{P}_n^{\tfrac{\nu+1}{2}}\left(\tfrac{A}{2\sqrt{B}}, \sinh^{-1}\tfrac{2\rho}{1-\rho^2}\right) & ,\mu > 2\sqrt{B} \end{cases} \tag{5.6}$$

The discrete energy spectrum requirement puts $B = (\mu/2)^2$ and gives

$$E_n = -\tfrac{\lambda^2}{2}\left(\tfrac{A}{\mu}+n+\tfrac{1}{2}\right)^2. \tag{5.7}$$

Therefore, the corresponding bound states wavefunctions are as follows

$$\psi_n(y) = A_n e^{-\lambda|\tfrac{A}{\mu}+n+\tfrac{1}{2}|y} \exp\left(-\sqrt{B}\,e^{-\lambda y}\right) L_n^{|\tfrac{2A}{\mu}+2n+1|}\left(2\sqrt{B}\,e^{-\lambda y}\right). \tag{5.8}$$

Case (2) $\alpha = (\nu+1)/2$:

$$V = \tfrac{\lambda^2}{2}\left[A e^{-\lambda r} + \left(\tfrac{\mu}{2}\right)^2 e^{-2\lambda r}\right]. \tag{5.9}$$

$$\frac{2}{\lambda^2}\langle \phi_m | H - E | \phi_n \rangle = \left[(2n+\nu+1)\left(n+\tfrac{\nu}{2}+1+\tfrac{A}{\mu}\right) - n - \left(\tfrac{\nu+1}{2}\right)^2 - \tfrac{2E}{\lambda^2}\right]\delta_{n,m}$$
$$-\left(n+\tfrac{\nu}{2}+\tfrac{A}{\mu}\right)\sqrt{n(n+\nu)}\,\delta_{n,m+1} - \left(n+\tfrac{\nu}{2}+1+\tfrac{A}{\mu}\right)\sqrt{(n+1)(n+\nu+1)}\,\delta_{n,m-1} \tag{5.10}$$

$$f_n(E) = \sqrt{\tfrac{\Gamma(n+\nu+1)}{\Gamma(n+1)}}\, S_n^{\tfrac{\nu+1}{2}}\left(z; \tfrac{\nu+1}{2}, \tfrac{A}{\mu}+\tfrac{1}{2}\right), \tag{5.11}$$

where $z = \sqrt{\left(\tfrac{\nu+1}{2}\right)^2 + 2E/\lambda^2}$. The solution of this case coincides with the findings in Refs. [16]. Moreover, we also obtain by diagonalization of (5.10) the following discrete energy spectrum

$$E_n = -\tfrac{\lambda^2}{2}\left(\tfrac{\nu+1}{2}\right)^2 = -\tfrac{\lambda^2}{2}\left(\tfrac{A}{\mu}+n+\tfrac{1}{2}\right)^2, \tag{5.12}$$

which is identical to (5.7) above.

## VI. THE S-WAVE HULTHÉN PROBLEM

The Hulthén potential [17], which is written as $V(r) = -\lambda \mathcal{Z} e^{-\lambda r}/(1-e^{-\lambda r})$ with $\lambda > 0$, is used as a model for a screened Coulomb potential, where $\lambda$ is the screening parameter. This is so, because for small $\lambda$ we can write the potential as $V(r) \approx -\tfrac{\mathcal{Z}}{r} e^{-\lambda r}$. The configuration space coordinate which is compatible with these kind of problems is $x = 1 - 2e^{-\lambda r}$. It maps real space into a bounded one. That is, $x \in [-1, +1]$ for $r \in [0, \infty]$. This problem belongs to the situation described by Eq. (1.7) with $x_\pm = \pm 1$. Since ${}_2F_1(-n, n+b; c; z)$ is proportional to the Jacobi polynomial $P_n^{(c-1, b-c)}(1-2z)$ [9], then the $L^2$ basis functions that satisfy the boundary conditions for this case could be written as

$$\phi_n(r) = A_n (1+x)^\alpha (1-x)^\beta P_n^{(\mu,\nu)}(x), \tag{6.1}$$

where $\alpha, \beta > 0$, $\mu, \nu > -1$ and the normalization constant is

$$A_n = \sqrt{\frac{\lambda(2n+\mu+\nu+1)}{2^{\mu+\nu+1}} \frac{\Gamma(n+1)\Gamma(n+\mu+\nu+1)}{\Gamma(n+\mu+1)\Gamma(n+\nu+1)}}. \tag{6.2}$$

Using the differential formulas of the Jacobi polynomials [Eqs. (A.8) and (A.9) in the Appendix], and $\frac{d}{dr} = \lambda(1-x)\frac{d}{dx}$, we can write

$$\frac{d^2\phi_n}{dr^2} = \lambda^2 \frac{1-x}{1+x} \left\{ \left[ -n\left(x + \frac{\nu-\mu}{2n+\mu+\nu}\right)\left(\frac{\mu-2\beta}{1-x} + \frac{2\alpha-\nu-1}{1+x}\right) - n(n+\mu+\nu+1) - \alpha(2\beta+1) \right.\right.$$
$$\left.\left. +\beta^2 \frac{1+x}{1-x} + \alpha(\alpha-1)\frac{1-x}{1+x} \right] \phi_n + 2 \frac{(n+\mu)(n+\nu)}{2n+\mu+\nu}\left(\frac{\mu-2\beta}{1-x} + \frac{2\alpha-\nu-1}{1+x}\right) \frac{A_n}{A_{n-1}} \phi_{n-1} \right\}. \tag{6.3}$$

Noting that $\int_0^\infty dr = \frac{1}{\lambda} \int_{-1}^{+1} \frac{dx}{1-x}$ and using the orthogonality relation for the Jacobi polynomials [Eq. (A.10) in the Appendix], we arrive at the following conclusions. First, this problem admits only S-wave ($\ell = 0$) exact solutions since the orbital term creates intractable non-tridiagonal representations. Second, the tridiagonal requirement on the action of the wave operator limits the possibilities to the following three:

(1) $\beta = \mu/2$, and $\alpha = (\nu+1)/2$   (6.4a)
(2) $\beta = \mu/2$, and $\alpha = 1 + \nu/2$   (6.4b)
(3) $\beta = (\mu+1)/2$, and $\alpha = (\nu+1)/2$   (6.4c)

The first possibility eliminates the $\phi_{n-1}$ term from Eq. (6.3), whereas the last two allow this term to contribute to the matrix elements above and below the diagonal. The calculation in the first possibility (6.4a) gives the following action of the S-wave Schrödinger operator on the basis:

$$\frac{2}{\lambda^2}(H-E)|\phi_n\rangle = \frac{1-x}{1+x}\left[ n(n+\mu+\nu+1) + \frac{1}{2}(\mu+1)(\nu+1) \right.$$
$$\left. -\frac{\mu^2}{4}\frac{1+x}{1-x} - \frac{\nu^2-1}{4}\frac{1-x}{1+x} + \frac{2}{\lambda^2}\frac{1+x}{1-x}(V-E) \right]|\phi_n\rangle. \tag{6.5}$$

Therefore, to obtain the tridiagonal representation, our choice of potential functions is limited to those that satisfy the following constraint:

$$\frac{2}{\lambda^2}\frac{1+x}{1-x}(V-E) = +\frac{\mu^2}{4}\frac{1+x}{1-x} + \frac{\nu^2-1}{4}\frac{1-x}{1+x} + \frac{2A}{\lambda^2} + \frac{B}{\lambda^2}(1\pm x), \tag{6.6}$$

where $A$ and $B$ are real potential parameters. The first two terms on the right hand side of the equation are necessary to cancel the contribution of the corresponding terms in Eq. (6.5) that destroy the tridiagonal structure. Equation (6.6) results in the following

$$E = -\frac{1}{2}(\mu\lambda/2)^2, \tag{6.7}$$

$$V = \frac{C}{(e^{\lambda r}-1)^2} + \frac{A}{e^{\lambda r}-1} + Be^{-\lambda r} \times \begin{cases} 1 \\ \frac{1}{e^{\lambda r}-1} \end{cases} \tag{6.8}$$

where $C = \frac{\lambda^2}{2}\frac{v^2-1}{4}$ and $E < 0$. The two alternatives in the last term of the potential correspond to the ± sign in Eq. (6.6). From now on, we will adopt the + sign making the last potential term in (6.8) pure exponential, $Be^{-\lambda r}$. However, the other choice could easily be obtained from this one by the parameter map: $B \to -B$, $A \to A + B$. The first two terms in $V(r)$ are the Hulthén potential and its square. After some manipulations, we obtain the following matrix representation of the wave operator

$$\frac{2}{\lambda^2}\langle\phi_m|H - E|\phi_n\rangle = \left[n(n+\mu+v+1) + \frac{2B}{\lambda^2}\frac{2n(n+\mu+v+1)+(\mu+v)(v+1)}{(2n+\mu+v)(2n+\mu+v+2)}\right.$$
$$\left.+\frac{1}{2}(\mu+1)(v+1) + \frac{2A}{\lambda^2}\right]\delta_{n,m} + \frac{2B/\lambda^2}{2n+\mu+v}\sqrt{\frac{n(n+\mu)(n+v)(n+\mu+v)}{(2n+\mu+v-1)(2n+\mu+v+1)}}\delta_{n,m+1} \qquad (6.9)$$
$$+\frac{2B/\lambda^2}{2n+\mu+v+2}\sqrt{\frac{(n+1)(n+\mu+1)(n+v+1)(n+\mu+v+1)}{(2n+\mu+v+1)(2n+\mu+v+3)}}\delta_{n,m-1},$$

where $\mu = \mu(E)$ as given by Eq. (6.7). The resulting three-term recursion relation (1.2) could be written in terms of polynomials defined by

$$P_n(E) = \frac{1}{\sqrt{2n+\mu+v+1}}\sqrt{\frac{\Gamma(n+\mu+1)\Gamma(n+v+1)}{\Gamma(n+1)\Gamma(n+\mu+v+1)}} f_n(E), \qquad (6.10)$$

in which case it reads

$$z P_n = \left[\gamma\left(n+\frac{\mu+v+1}{2}\right)^2 + \frac{2n(n+\mu+v+1)+(\mu+v)(v+1)}{(2n+\mu+v)(2n+\mu+v+2)}\right]P_n$$
$$+\frac{(n+\mu)(n+v)}{(2n+\mu+v)(2n+\mu+v+1)}P_{n-1} + \frac{(n+1)(n+\mu+v+1)}{(2n+\mu+v+1)(2n+\mu+v+2)}P_{n+1} \qquad (6.11)$$

where $\gamma = \lambda^2/2B$ and $z(E) = (C - A - E)/B$. We are not aware of any known three-parameter orthogonal polynomials that satisfy the above recursion relation. However, comparing it to the recursion (A.6) in the Appendix suggests that these polynomials could be considered as deformations of the Jacobi polynomials with $\gamma$ being the deformation parameter ($\gamma = 0$ corresponds to the Jacobi polynomial). Pursuing the analysis of these polynomials would be too mathematical and inappropriate for the present setting. Nonetheless, we find it pressing to make the following remark. For large values of the index $n$ the $\gamma$ term in the recursion (6.11) goes like $n^2$ whereas the rest of the terms go like $n^0$. Therefore, to obtain reasonable and meaningful numerical results, $\gamma$ should be taken very small (i.e., $\lambda^2 \ll B$). Now back to the matrix representation (6.9) of the wave operator. The discrete energy spectrum is easily obtained by diagonalizing this representation which requires that $B = 0$ and

$$n(n+\mu+v+1) + \frac{1}{2}(\mu+1)(v+1) + 2A/\lambda^2 = 0, \qquad (6.12)$$

giving the following discrete spectrum for $n = 0, 1, 2, ...$

$$E_n = -\frac{\lambda^2}{2}\left(\frac{\mu}{2}\right)^2 = -\frac{\lambda^2}{8}\left[n+\frac{v+1}{2} + \frac{2(A-C)/\lambda^2}{n+\frac{v+1}{2}}\right]^2. \qquad (6.13)$$

The corresponding bound states could be written as follows

$$\psi_n(r) = A_n e^{-\lambda\mu_n r/2}(1-e^{-\lambda r})^{\frac{v+1}{2}} P_n^{(\mu_n, v)}(1-2e^{-\lambda r}), \qquad (6.14)$$

where $\mu_n = \frac{2}{\lambda}\sqrt{-2E_n}$ and $v = \sqrt{1+8C/\lambda^2}$. These findings, for the special case where $B = 0$ in the potential function (6.8), agree with the results obtained in Refs. [18].

Repeating the same analysis for the second possibility (6.4b) and investigating the tridiagonal structure of the resulting action of the wave operator on the basis we conclude the following. First, the parameter $\mu$ is related to the energy by Eq. (6.7) – the same way as in the first case. Second, the potential is required to take the following functional form

$$V = A\frac{1-x}{1+x} + 2B\frac{1-x}{(1+x)^2} = \frac{A}{e^{\lambda r}-1} + \frac{Be^{\lambda r}}{(e^{\lambda r}-1)^2} = \frac{A+B}{e^{\lambda r}-1} + \frac{B}{(e^{\lambda r}-1)^2} \qquad (6.15)$$

The matrix representation of the S-wave Schrödinger operator is obtained as

$$\begin{aligned}\frac{1}{\lambda^2}\langle\phi_m|H-E|\phi_n\rangle = & \\ & \left\{-z - \frac{n(n+\mu)}{2n+\mu+\nu} + \frac{2n(n+\mu+\nu+1)+(\mu+\nu)(\nu+1)}{(2n+\mu+\nu)(2n+\mu+\nu+2)}\left[\left(n+\tfrac{\mu+\nu}{2}+1\right)^2+\gamma\right]\right\}\delta_{n,m} \\ & + \frac{1}{2n+\mu+\nu}\sqrt{\frac{n(n+\mu)(n+\nu)(n+\mu+\nu)}{(2n+\mu+\nu-1)(2n+\mu+\nu+1)}}\left[\left(n+\tfrac{\mu+\nu}{2}\right)^2+\gamma\right]\delta_{n,m+1} \\ & + \frac{1}{2n+\mu+\nu+2}\sqrt{\frac{(n+1)(n+\mu+1)(n+\nu+1)(n+\mu+\nu+1)}{(2n+\mu+\nu+1)(2n+\mu+\nu+3)}}\left[\left(n+\tfrac{\mu+\nu}{2}+1\right)^2+\gamma\right]\delta_{n,m-1}\end{aligned} \qquad (6.16)$$

where $z = \left(\frac{\nu+1}{2}\right)^2 - 2B/\lambda^2 - 1/4$ and $\gamma = 2(E+A)/\lambda^2$. The resulting recursion relation in terms of the polynomials defined by Eq. (6.10) above reads as follows

$$\begin{aligned}zP_n = & \left\{-\frac{n(n+\mu)}{2n+\mu+\nu} + \frac{2n(n+\mu+\nu+1)+(\mu+\nu)(\nu+1)}{(2n+\mu+\nu)(2n+\mu+\nu+2)}\left[\left(n+\tfrac{\mu+\nu}{2}+1\right)^2+\gamma\right]\right\}P_n \\ & + \frac{(n+\mu)(n+\nu)}{(2n+\mu+\nu)(2n+\mu+\nu+1)}\left[\left(n+\tfrac{\mu+\nu}{2}\right)^2+\gamma\right]P_{n-1} \\ & + \frac{(n+1)(n+\mu+\nu+1)}{(2n+\mu+\nu+1)(2n+\mu+\nu+2)}\left[\left(n+\tfrac{\mu+\nu}{2}+1\right)^2+\gamma\right]P_{n+1}\end{aligned} \qquad (6.17)$$

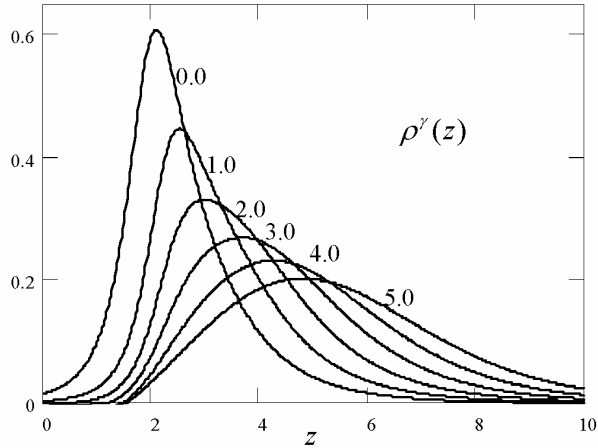

**Fig. 1**: A graph of the density (weight) function $\rho^\gamma(z)$ associated with the orthogonal polynomials satisfying the three-term recursion relation (6.17). The "Dispersion Correction" method developed in Ref. [19] was used to generate this plot using the recursion coefficients $\{a_n,b_n\}_{n=0}^{50}$ in Eq. (6.16) with $\mu = 1.0$, $\nu = 1.5$ and for several values of the parameter $\gamma$ as shown on the traces.

The orthogonal polynomials defined by this recursion relation are, to the best of our knowledge, not seen before. The mathematical analysis of this recursion relation and the corresponding polynomials will not be carried out here. However, we find it appropriate and physically sufficient to give a fairly accurate graphical representation of the density (weight) function $\rho^\gamma(z)$ associated with the orthogonality of these polynomials. We use one of three numerical methods developed in Ref. [19] to obtain a good approximation of the density function associated with a finite tridiagonal Hamiltonian matrix. Figure (1) shows $\rho^\gamma(z)$ for a given value of $\mu$ and $\nu$ and for several choices of the parameter $\gamma$. The discrete energy spectrum is obtained (by diagonalization) from Eq. (6.16) as

$$E_n = -\frac{\lambda^2}{8}\left[n+\frac{\nu+1}{2}+1+\frac{2A/\lambda^2}{n+\frac{\nu+1}{2}+1}\right]^2, \quad n=0,1,2,... \tag{6.18}$$

where $\frac{\nu+1}{2} = |D/\lambda|$ and $D$ is defined by $D^2 = 2B + \lambda^2/4$ where $B > -\lambda^2/8$.

We leave it to the interested reader to find the recursion relation for the third possibility (6.4c) and to verify the following results:

$$V = \frac{C}{(e^{\lambda r}-1)^2} + \frac{A}{e^{\lambda r}-1} \tag{6.19}$$

where $C = \frac{\lambda^2}{2}\frac{\nu^2-1}{4}$ and $A$ is a real potential parameter. Moreover,

$$\frac{1}{\lambda^2}\langle\phi_m|H-E|\phi_n\rangle =$$
$$\left\{-z-\frac{n(n+\nu)}{2n+\mu+\nu}+\frac{2n(n+\mu+\nu+1)+(\mu+\nu)(\mu+1)}{(2n+\mu+\nu)(2n+\mu+\nu+2)}\left[\left(n+\frac{\mu+\nu}{2}+1\right)^2+\gamma\right]\right\}\delta_{n,m}$$
$$-\frac{1}{2n+\mu+\nu}\sqrt{\frac{n(n+\mu)(n+\nu)(n+\mu+\nu)}{(2n+\mu+\nu-1)(2n+\mu+\nu+1)}}\left[\left(n+\frac{\mu+\nu}{2}\right)^2+\gamma\right]\delta_{n,m+1}$$
$$-\frac{1}{2n+\mu+\nu+2}\sqrt{\frac{(n+1)(n+\mu+1)(n+\nu+1)(n+\mu+\nu+1)}{(2n+\mu+\nu+1)(2n+\mu+\nu+3)}}\left[\left(n+\frac{\mu+\nu}{2}+1\right)^2+\gamma\right]\delta_{n,m-1} \tag{6.20}$$

Aside from some sign changes and the exchange $\mu \leftrightarrow \nu$, this is the same as Eq. (6.16) above. However, here we have $z = \frac{2E}{\lambda^2}+\left(\frac{\mu+1}{2}\right)^2$ and $\gamma = 2(E+A-C)/\lambda^2$. The discrete energy spectrum is obtained as

$$E_n = -\frac{\lambda^2}{2}\left(\frac{\mu+1}{2}\right)^2 = -\frac{\lambda^2}{8}\left[n+\frac{\nu+1}{2}+\frac{2(A-C)/\lambda^2}{n+\frac{\nu+1}{2}}\right]^2 \tag{6.21}$$

VII. ROSEN-MORSE TYPE POTENTIALS

The basis functions for the one-dimensional problem associated with this potential could be written as follows

$$\phi_n(x) = A_n(1+x)^\alpha(1-x)^\beta P_n^{(\mu,\nu)}(x), \tag{7.1}$$

where $x = \tanh(\lambda y)$ and $y \in \Re^1$. The parameters $\alpha, \beta, \mu, \nu, \lambda$ are real with $\alpha, \beta$ and $\lambda$ positive. The normalization constant $A_n$ is given by Eq. (6.2). Analytic solutions of this problem are obtainable for three cases where the parameters are related as: $(\alpha,\beta) = \left(\frac{\nu}{2},\frac{\mu}{2}\right), \left(\frac{\nu}{2},\frac{\mu+1}{2}\right),$ or $\left(\frac{\nu+1}{2},\frac{\mu}{2}\right)$. As an example, we consider only the first case where the potential function assumes the following form which is compatible with the tridiagonal representation

$$V = C\tanh(\lambda y) + \frac{A}{\cosh(\lambda y)^2} \pm B\frac{\tanh(\lambda y)}{\cosh(\lambda y)^2}, \qquad (7.2)$$

where $C = (\lambda\mu/2)^2 - (\lambda\nu/2)^2$. We consider, in what follows, the potential with the top + sign in (7.2). The tridiagonal representation of the wave operator becomes

$$\frac{2}{\lambda^2}\langle\phi_m|H-E|\phi_n\rangle = \left[2\frac{A-B}{\lambda^2} - \frac{1}{4} + \left(n + \frac{\mu+\nu+1}{2}\right)^2 + \frac{4B}{\lambda^2}\frac{2n(n+\mu+\nu+1)+(\mu+\nu)(\nu+1)}{(2n+\mu+\nu)(2n+\mu+\nu+2)}\right]\delta_{n,m}$$
$$+ \frac{4B/\lambda^2}{2n+\mu+\nu}\sqrt{\frac{n(n+\mu)(n+\nu)(n+\mu+\nu)}{(2n+\mu+\nu-1)(2n+\mu+\nu+1)}}\delta_{n,m+1} \qquad (7.3)$$
$$+ \frac{4B/\lambda^2}{2n+\mu+\nu+2}\sqrt{\frac{(n+1)(n+\mu+1)(n+\nu+1)(n+\mu+\nu+1)}{(2n+\mu+\nu+1)(2n+\mu+\nu+3)}}\delta_{n,m-1}$$

The resulting recursion relation is similar to (6.11). The discrete energy spectrum is obtainable only for $B = 0$ which corresponds to the hyperbolic Rosen-Morse potential [20]. In this case we obtain:

$$E_n = -(\lambda\mu/2)^2 - (\lambda\nu/2)^2 = -\frac{\lambda^2}{2}\left[\left(n + \tfrac{1}{2} - |D/\lambda|\right)^2 + \left(\frac{C}{\lambda^2}\right)^2\left(n + \tfrac{1}{2} - |D/\lambda|\right)^{-2}\right] \qquad (7.4)$$

where $D$ is defined by $D^2 = -2A + \lambda^2/4$ and $A < \lambda^2/8$. The corresponding bound states wavefunctions are

$$\psi_n(y) = A_n(1 + \tanh\lambda y)^{\nu_n/2}(1 - \tanh\lambda y)^{\mu_n/2}P_n^{(\mu_n,\nu_n)}(\tanh\lambda y), \qquad (7.5)$$

where $\mu_n = \frac{1}{\lambda}\sqrt{-2(E_n - C)}$ and $\nu_n = \frac{1}{\lambda}\sqrt{-2(E_n + C)}$.

Finally, we note that the examples presented in this work do not exhaust all possible potentials in this larger class of analytically solvable systems. Moreover, it might be possible that this approach could be extended to the study of quasi exactly and conditionally exactly solvable problems. In addition, the relativistic extension of this development is also possible. In fact, the Dirac-Coulomb and Dirac-Morse problems have already been worked out [21].

## APPENDIX

The following are useful formulas and relations satisfied by the orthogonal polynomials that are relevant to the development carried out in this work. They are found in most books on orthogonal polynomials [9]. We list them here for ease of reference.

(1) The Laguerre polynomials $L_n^\nu(x)$, where $\nu > -1$:

$$xL_n^\nu = (2n+\nu+1)L_n^\nu - (n+\nu)L_{n-1}^\nu - (n+1)L_{n+1}^\nu \qquad (A.1)$$

$$L_n^\nu(x) = \frac{\Gamma(n+\nu+1)}{\Gamma(n+1)\Gamma(\nu+1)}{}_1F_1(-n;\nu+1;x) \qquad (A.2)$$

$$\left[x\frac{d^2}{dx^2} + (\nu+1-x)\frac{d}{dx} + n\right]L_n^\nu(x) = 0 \qquad (A.3)$$

$$x\frac{d}{dx}L_n^\nu = nL_n^\nu - (n+\nu)L_{n-1}^\nu \qquad (A.4)$$

$$\int_0^\infty x^\nu e^{-x}L_n^\nu(x)L_m^\nu(x)dx = \frac{\Gamma(n+\nu+1)}{\Gamma(n+1)}\delta_{nm} \qquad (A.5)$$

(2) The Jacobi polynomials $P_n^{(\mu,\nu)}(x)$, where $\mu > -1, \nu > -1$:

$$\left(\frac{1\pm x}{2}\right)P_n^{(\mu,\nu)} = \frac{2n(n+\mu+\nu+1)+(\mu+\nu)(\frac{\mu+\nu}{2}\pm\frac{\nu-\mu}{2}+1)}{(2n+\mu+\nu)(2n+\mu+\nu+2)} P_n^{(\mu,\nu)}$$
$$\pm \frac{(n+\mu)(n+\nu)}{(2n+\mu+\nu)(2n+\mu+\nu+1)} P_{n-1}^{(\mu,\nu)} \pm \frac{(n+1)(n+\mu+\nu+1)}{(2n+\mu+\nu+1)(2n+\mu+\nu+2)} P_{n+1}^{(\mu,\nu)}$$
(A.6)

$$P_n^{(\mu,\nu)}(x) = \frac{\Gamma(n+\mu+1)}{\Gamma(n+1)\Gamma(\mu+1)}\, {}_2F_1(-n,n+\mu+\nu+1;\mu+1;\tfrac{1-x}{2}) = (-)^n P_n^{(\nu,\mu)}(-x) \quad (A.7)$$

$$\left\{(1-x^2)\frac{d^2}{dx^2} - \left[(\mu+\nu+2)x+\mu-\nu\right]\frac{d}{dx} + n(n+\mu+\nu+1)\right\}P_n^{(\mu,\nu)}(x) = 0 \quad (A.8)$$

$$(1-x^2)\frac{d}{dx}P_n^{(\mu,\nu)} = -n\left(x+\frac{\nu-\mu}{2n+\mu+\nu}\right)P_n^{(\mu,\nu)} + 2\frac{(n+\mu)(n+\nu)}{2n+\mu+\nu}P_{n-1}^{(\mu,\nu)} \quad (A.9)$$

$$\int_{-1}^{+1}(1-x)^\mu(1+x)^\nu P_n^{(\mu,\nu)}(x)P_m^{(\mu,\nu)}(x)dx = \frac{2^{\mu+\nu+1}}{2n+\mu+\nu+1}\frac{\Gamma(n+\mu+1)\Gamma(n+\nu+1)}{\Gamma(n+1)\Gamma(n+\mu+\nu+1)}\delta_{nm} \quad (A.10)$$

(3) The Pollaczek polynomials $P_n^\mu(x,\theta)$, where $\mu > 0$ and $0 < \theta < \pi$:

$$2[(n+\mu)\cos\theta + x\sin\theta]P_n^\mu - (n+2\mu-1)P_{n-1}^\mu - (n+1)P_{n+1}^\mu = 0 \quad (A.11)$$

$$P_n^\mu(x,\theta) = \frac{\Gamma(n+2\mu)}{\Gamma(n+1)\Gamma(2\mu)} e^{in\theta}\, {}_2F_1(-n,\mu+ix;2\mu;1-e^{-2i\theta}) \quad (A.12)$$

$$\int_{-\infty}^{+\infty} \rho^\mu(x,\theta) P_n^\mu(x,\theta) P_m^\mu(x,\theta) dx = \frac{\Gamma(n+2\mu)}{\Gamma(n+1)}\delta_{nm} \quad (A.13)$$

where $\rho^\mu(x,\theta) = \frac{1}{2\pi}(2\sin\theta)^{2\mu} e^{(2\theta-\pi)x}|\Gamma(\mu+ix)|^2$.

(4) The continuous dual Hahn polynomials $S_n^\mu(x;a,b)$, where $x^2 > 0$ and $\mu$, $a$, $b$ are positive except for a pair of complex conjugates with positive real parts:

$$x^2 S_n^\mu = \left[(n+\mu+a)(n+\mu+b)+n(n+a+b-1)-\mu^2\right]S_n^\mu$$
$$-n(n+a+b-1)S_{n-1}^\mu - (n+\mu+a)(n+\mu+b)S_{n+1}^\mu$$
(A.14)

$$S_n^\mu(x;a,b) = {}_3F_2\left(\begin{matrix}-n,\mu+ix,\mu-ix\\ \mu+a,\mu+b\end{matrix}\bigg|1\right) \quad (A.15)$$

$$\int_0^\infty \rho^\mu(x) S_n^\mu(x;a,b) S_m^\mu(x;a,b) dx = \frac{\Gamma(n+1)\Gamma(n+a+b)}{\Gamma(n+\mu+a)\Gamma(n+\mu+b)}\delta_{nm} \quad (A.16)$$

where $\rho^\mu(x) = \frac{1}{2\pi}\left|\frac{\Gamma(\mu+ix)\Gamma(a+ix)\Gamma(b+ix)}{\Gamma(\mu+a)\Gamma(\mu+b)\Gamma(2ix)}\right|^2$.